\documentclass[conference]{IEEEtran}

\usepackage{amsmath}
\usepackage{graphicx}
\usepackage{subfig}

\usepackage{fancyhdr}
\begin{document}

\title{Parameter Setting for Quantum Annealers}

\author{\IEEEauthorblockN{Kristen L. Pudenz}
\IEEEauthorblockA{Lockheed Martin Aeronautics
Fort Worth, Texas\\
Email: kristen.l.pudenz@lmco.com}}

\maketitle

\begin{abstract}
We develop and apply several strategies for setting physical parameters on quantum annealers for application problems that do not fit natively on the hardware graph. The strategies are tested with a culled random set of mixed satisfiability problems, yielding results that generalize to guidelines regarding which parameter setting strategies to use for different classes of problems, and how to choose other necessary hardware quantities as well. Alternate methods of changing the hardware implementation of an application problem are also considered and their utility discussed.
\end{abstract}

\section{Introduction and Motivation}
At this point in the development of practical quantum annealing devices, there has been a great deal published regarding specialized benchmarking problems designed for existing hardware \cite{hen2015probing,king2015performance,ronnow2014defining,boixo2014evidence,King2014}. Progress has also been made in the development of potential applications that are suitable for quantum annealers but not fundamentally hardware specific \cite{hen2016quantum,rieffel2015case,venturelli2015jobshop,perdomo2015fault,Zick2015,adachi2015application,neven2009nips,pudenz2013quantum}. Although some work exists \cite{Perdomo-Ortiz2015,Perdomo-Ortiz2015a,venturelli2015quantum}, less attention has been paid in the published literature to the difficult and important work of transforming these more general applications into specific hardware implementations.

The first step in the transformation, embedding, while crucial to the success of quantum annealing (QA), is not the primary focus here. The leading embedding algorithm is that which has been developed by D-Wave Systems for their own hardware \cite{cai2014practical}, which we will use here without comment. This algorithm produces an \textit{embedding} which translates the $n$ \textit{logical qubits} $s_i$ that make up the original \textit{logical problem} $H_{logical}$ 

\begin{equation}
H_{logical} = \sum_{i=1}^{n} h_i s_i + \sum_{(i,j)\in E_{l}} J_{ij} s_i s_j
\label{eq_H_logical}
\end{equation}

into sets of $K_i$ \textit{physical qubits} $s_{i,k}$ that make up the \textit{physical problem} $H_{physical}$.

\begin{align}
H_{physical} = & s'As = \sum_{i=1}^{n} \sum_{k=1}^{K_i} h_{i,k} s_{i,k} + \\ & \sum_{(i,j)\in E_{l}} \sum_{(i,k;j,m)\in E_p} J_{i,k;j,m} s_{i,k} s_{j,m} + H_{chain}
\label{eq_H_phys}
\end{align}

The logical problem may define a \textit{logical bias} $h_i$ for each individual logical qubit, which will be represented by one or more \textit{physical bias} $h_{i,k}$ terms on the relevant physical qubits. Two logical qubits may be connected by a \textit{logical coupler} $J_{ij}$ in the logical problem, which must be represented by one or more \textit{physical couplers} $J_{i,k;j,m}$ in the physical problem. Finally, the physical qubits comprising each logical qubit must be kept consistent, and are therefore subject to \textit{chain couplings} $c$ connecting them.

\begin{equation}
H_{chain} = \sum_{i=1}^{n} \sum_{(i,k;i,m)\in E_p} c s_{i,k} s_{i,m}
\end{equation}

In this notation, $n$ is the number of logical qubits, $N$ is the number of physical qubits, $E_l$ is the logical edge set of the problem and $E_p$ is the physical edge set that exists in the hardware.

Even with an embedding provided, we are free to choose the physical biases, physical couplings, and chain couplings that transform a logical problem into a physical problem, subject to certain constraints. This process is called \textit{parameter setting}, and is addressed in Section \ref{ps_strategies}. Section \ref{decoding} covers \textit{decoding}, defined for our purposes as the process of transforming a \textit{physical result} back into a \textit{logical result} that is a candidate solution to the original logical problem. After defining our main procedural tools, we describe our test problem set of mixed satisfiability instances and commercial quantum annealing equipment in Sections \ref{mixed_sat} and \ref{equipment}. We then apply our parameter setting methods to the problem set, developing recommendations for choosing a favorable approach based on experimental data in Sections \ref{results} and \ref{conclusions}.

\section{Parameter Setting Strategies}
\label{ps_strategies}
The most important consideration in parameter setting is the preservation of the ratio between terms in the logical problem. Biases and couplings may be rescaled to suit the hardware parameters or the number and distribution of available qubits, but the relationship of each logical problem term (i.e. the sum of the associated physical problem terms) to each other term must remain consistent. The choice of chain coupling will in turn be affected by the physical problem terms being used.

\subsection{Single Device Programming}
The simplest parameter setting strategy is to choose one physical device on-chip to represent each logical problem term. Out of $K_i$ physical qubits in logical qubit $i$, one will be programmed with the full logical bias, so $h_{i,k} = h_i$ for the selected $k$ and $h_{i,k} = 0$ otherwise. If there is more than one physical coupler representing a logical coupler, only one is selected and programmed with the logical coupling, so one $J_{i,k;j,m} = J_{ij}$ and the rest are $0$. We choose the physical qubit with the most physical problem couplings attached to it, and the first available physical coupler (this choice is not as crucial because embedding algorithms often struggle to provide one physical coupler per logical coupler). This strategy has been observed to work best with chain couplings $c$ less than or equal to the magnitude of the largest physical problem term.

\subsection{Even Distribution}
Another straightforward strategy is to distribute logical problem terms evenly over the number of physical devices available to represent them. For a logical qubit with $K_i$ physical qubits, $h_{i,k} = h_i/K_i$ for all $k$, unless it falls below the hardware resolution $h_{min}$. In this case, we rank the physical qubits $s_{i,k} \in s_i$ in order of decreasing number of adjacent, active physical couplers. The first $| h_i/h_{min} |$ physical qubits are assigned $h_{i,k} = h_{min} \text{sign}(h_i)$, the next on the list gets the remainder, if any, from $h_i/h_{min}$, and the rest are assigned $h_{i,k} = 0$. The same procedure is followed for the couplers, although there are usually few of them and they cannot be ordered in the same way because every coupler touches exactly two qubits. The even distribution was generally found to be optimal when chain coupling exceeded the magnitude of the largest physical problem term, with this combination being the best overall strategy.

\subsection{Weighted Distribution}
\label{sec_weighted}
The idea that the physical qubits which are coupled to members of other logical groups are the most crucial leads to a weighted distribution of physical problem terms. We first assign a weight to each qubit, $w_{i,k} = d_{i,k}/D_{i}$, where $d_{i,k}$ is the number of active physical couplers attached to qubit $k$ of logical group $i$ and $D_i$ is the total number of active physical couplers for logical qubit $i$ (excluding couplers used in $H_{chain}$). The physical bias on each physical qubit is then $h_{i,k} = h_i w_{i,k}$. If $h_{i,k} < h_{min}$, $h_{i,k}$ is set to $0$ and its value is distributed among the physical qubits that made the cutoff.

A variant of this strategy, which we term the weighted regularized distribution, assigns a base bias of magnitude $h_{min}$ to each physical qubit, then distributes any extra bias according to the weight system; $h_{i,k} = h_{min} \text{sign}(h_i) + h_{remainder} w_{i,k}$, where $h_{remainder} = h_i - K_i h_{min} \text{sign}(h_i)$.The weighted regularized distribution results fell somewhere between those from the weighted and even strategies, more closely tracking the even strategy.

\section{Decoding Strategies}
\label{decoding}
After a QA problem has been implemented and annealed on hardware, the result must be transformed back into the solution space of the logical problem. We term this process decoding. For this work, several decoding strategies (single qubit, majority vote, and weighted majority vote) were used in concert.

\subsection{Single Qubit}
The simplest decoding strategy is to take the readout from the physical qubit within each logical qubit with the highest weight $w_{i,k}$ as described in section \ref{sec_weighted} to be the value of logical qubit $i$. This strategy never produces an indeterminate outcome, but also discards any information that may be contained in the remaining physical qubits.

\subsection{Majority Vote}
A more popular decoding strategy is to take a majority vote over the physical qubits to determine each logical qubit value. Ties in the majority vote are resolved by choosing one of the two possible values at random. The majority vote can be simple or it may take into account the weight $w_{i,k}$ of each physical qubit and give those with higher weights more value. The simple majority vote result is $v_i = \text{sign}(\sum_{k=1}^{K_i} s_{i,k})$ and the weighted majority vote result is $v_i = \text{sign}(\sum_{k=1}^{K_i} w_{i,k} s_{i,k})$, where the readout values $s_{i,k} \in \{-1,1\}$.

\section{The Mixed SAT Problem Set}
\label{mixed_sat}
We studied parameter setting on mixed satisfiability (SAT) problems. SAT problems are a core variety of optimization problems that underlie a variety of applications and are important enough to motivate the development of an ecosystem of specialized classical solvers \cite{hoos2000satllb,toda2015implementing}. Mixed SAT problems specifically have fewer constraints on their structure than other classes of SAT, making them a useful general class to study. Whereas many SAT classes are defined by the number of variables in the individual clauses that together constitute the larger problem, two clauses in a mixed SAT problem need not involve the same number of variables, opening up representations of naturally varied constraints. Another important feature of this problem class is that, unlike many problem sets that have been constructed for study with quantum annealers to date \cite{hen2015probing,king2015performance,ronnow2014defining,boixo2014evidence,King2014}, but like the general class of potential optimization applications, it does not embed natively on the hardware graph, making parameter setting critically important for performance. These problems also have multiple satisfying answers, a case not usually considered by classical SAT solvers due to the difficulty of finding all solutions, leaving room for progress on sampling over the solution space for problems with many satisfying assignments.

Mixed SAT problems involving $n\in\{10,20,30\}$ binary variables were generated by randomly choosing $\alpha=\{10, 20, 30, 40, 50\}$ clauses of bounded length. The random problem instance set was downselected twice, first to limit the number of solutions to fewer than a million (according to the results of a classical all-solutions SAT solver \cite{toda2015implementing}), then to instances for which we were able to find an embedding on hardware. Practically speaking, this meant that the $n\in\{10, 20\}$ instances were implemented on the smaller $504$ qubit D-Wave Two (DW2) chip used for this work, while the $n=30$ instances were studied using a larger $1098$ qubit D-Wave Two X (DW2X) processor. The resulting problem set had $927$ instances for the DW2, and 123 for the DW2X.

\section{Equipment}
\label{equipment}
Quantum annealing optimization experiments referenced here were performed using two commercial quantum annealing chips manufactured by D-Wave Systems. The DW2 generation chip was designed with $512$ qubits and yielded $504$ working qubits; the DW2X was designed with $1152$ qubits and yielded $1098$ working. This class of quantum annealing processor uses superconducting quantum interference devices to create a quantum environment for solving Ising spin glass problems of the form described in Equation \ref{eq_H_logical} and is extensively described in the literature \cite{johnson2011quantum,lanting2014entanglement,bunyk2014architectural}.

\section{Results}
\label{results}

\subsection{Chain Coupling Selection}
The first parameter to set is the magnitude of the chain coupling $c$, which ties the logical qubits into internally consistent units. The choice of this seemingly straightforward parameter can have crucial impacts on the time-dependent energy spectrum of the annealing process. If the chain coupling is too weak, logical qubits break into domains of physical qubits with opposing spin orientations, failing to act as a single variable. If it is too strong, it can overwhelm the physical problem terms $h_{i,k}$ and $J_{i,k;j,m}$, possibly even pushing them below the precision threshold of the devices on-chip (physical bias and coupling terms are subject to noise; those that are set too close together may in fact cross \cite{king2015performance,King2014}). This is an important point because the chain couplings that are necessary for internally consistent logical qubits are often greater than $1$, if we define the scaling of the physical problem such that the maximum physical problem term $max(h_{i,k}, J_{i,k;j,m}) = 1$.

\begin{figure}
\centering
\includegraphics[width=2.75in, angle=270]{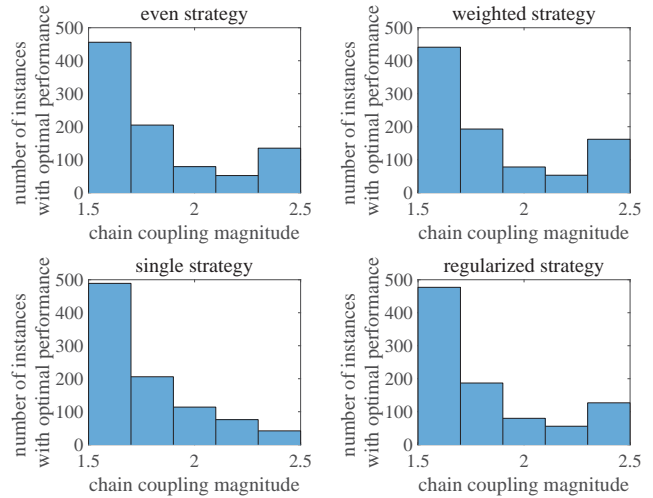}
\caption{Histogram of number of DW2 instances exhibiting their highest success probability at each chain coupling. One panel for each parameter setting strategy. For all strategies, a majority of instances had optimal performance at $c=1.6$.}
\label{DW2_hist}
\end{figure}

\begin{figure}
\centering
\includegraphics[width=2.75in]{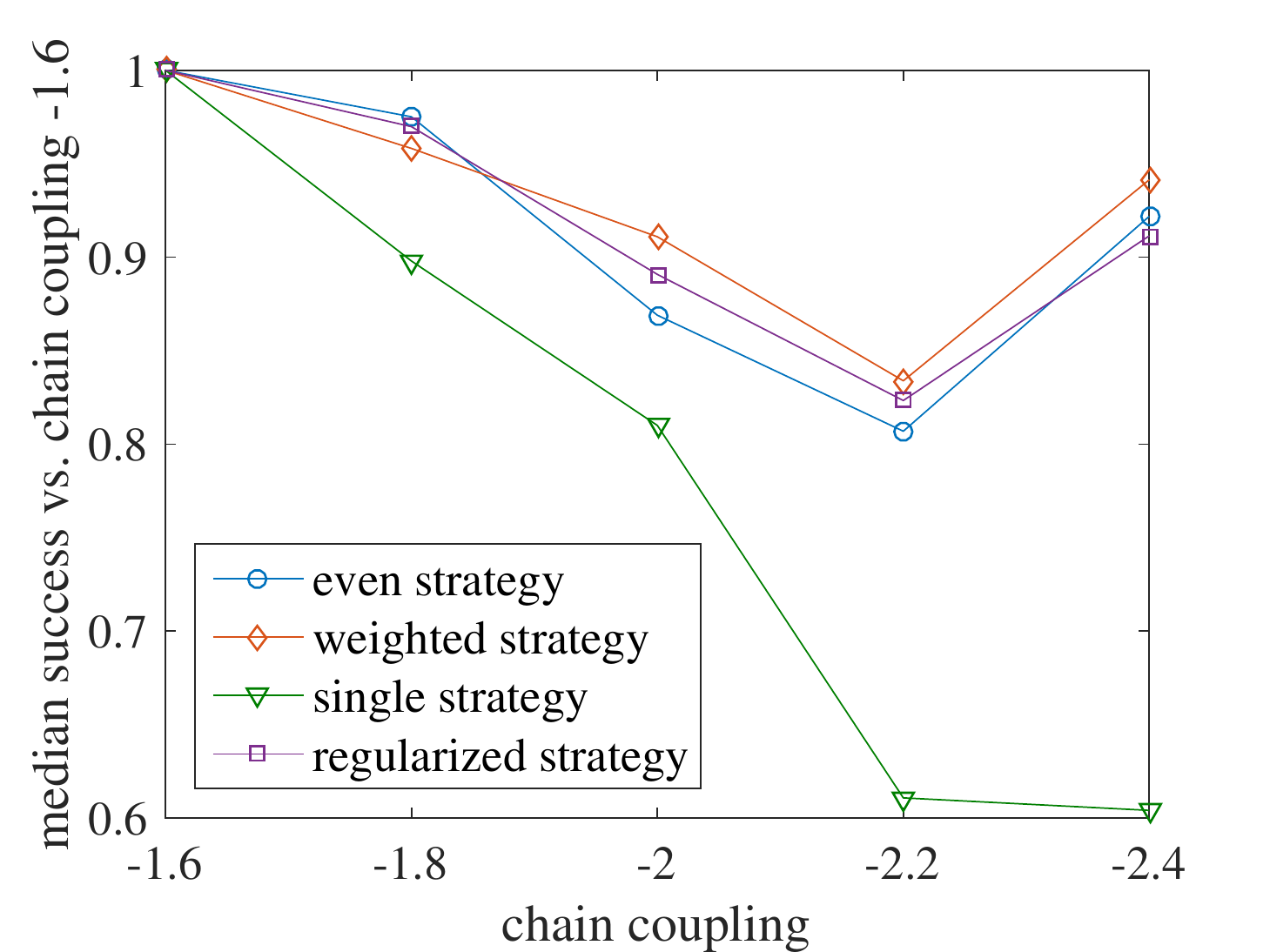}
\caption{Median success probability over the DW2 problem set as compared to success probability at $c=1.6$. Plotted here is the median value of $P_{success,i,c}/P_{success,i,1.6}$ over the problem set. All parameter setting strategies perform best for $c=1.6$, with the single strategy exhibiting a particularly sharp decline in success probability as the chain coupling increases.}
\label{DW2_median_success_vs_chain}
\end{figure}

Our mixed SAT problem set was tested with chain couplings $c\in{1.6,1.8,2.0,2.2,2.4}$, a range motivated by an earlier study of 3-SAT problems which showed the most instances were solved at $c=2.0$. Each problem instance has its own peak in success probability (the observed probability that a single annealing run will result in an answer satisfying the mixed SAT formula) and number of unique answers observed. In order to treat the problem set as a class, we examined two figures of merit for performance: the number of instances with optimal performance at a given chain coupling, and the median success probability ratio over the instance set for all problems. Because some instances are harder than others, we used a ratio normalizing the success probability of each problem at a given chain coupling to its performance at $c=1.6$, therefore differentiating the performance of the chain couplings regardless of instance hardness. The results for the DW2 ($n\in\{10,20\}$) problem set can be seen in Figures \ref{DW2_hist} and \ref{DW2_median_success_vs_chain}, and indicate an optimal chain coupling of $c=1.6$ over the parameter space studied.

\begin{figure}
\centering
\includegraphics[width=2.75in, angle=270]{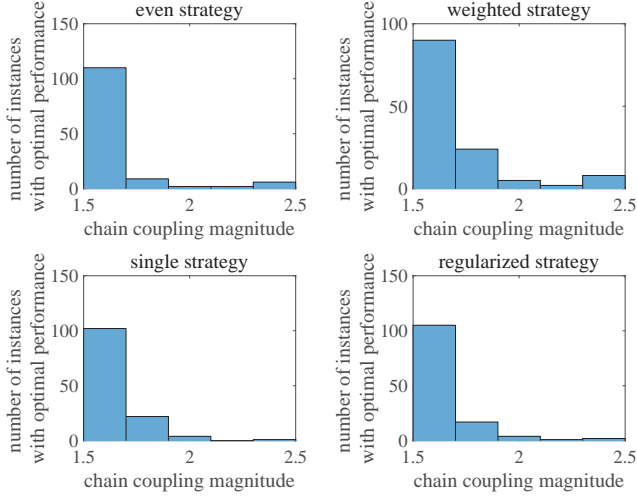}
\caption{Histogram of number of DW2X instances exhibiting their highest success probability at each chain coupling. One panel for each parameter setting strategy. Again, a majority of instances had optimal performance at $c=1.6$.}
\label{DW2X_hist}
\end{figure}

\begin{figure}
\centering
\includegraphics[width=2.75in]{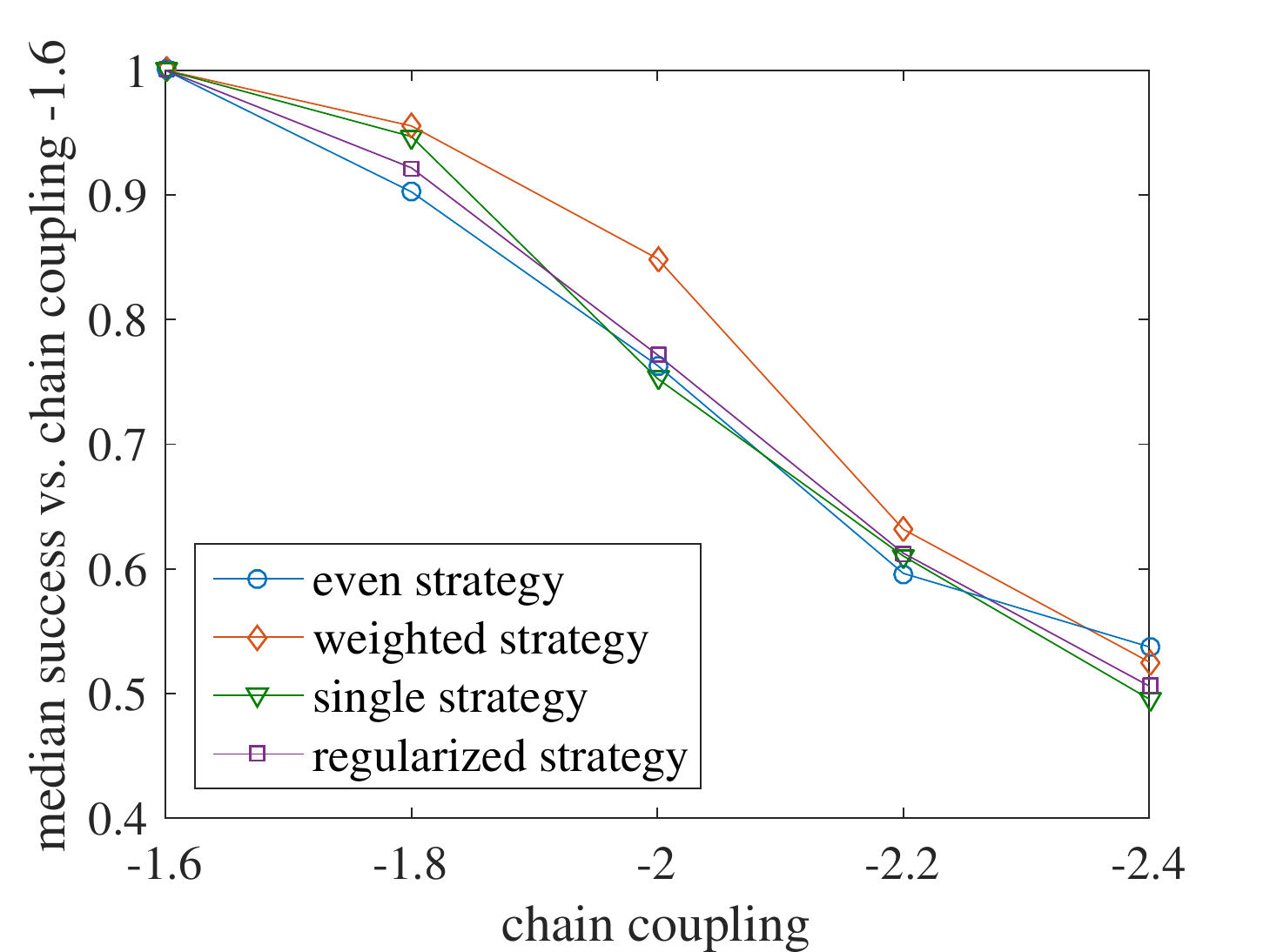}
\caption{Median success probability over the DW2X problem set as compared to success probability at $c=1.6$. Plotted here is the median value of $P_{success,i,c}/P_{success,i,1.6}$ over the problem set. All parameter setting strategies perform best for $c=1.6$, and decline more quickly as the chain coupling increases than was observed for the DW2 problem set.}
\label{DW2X_median_success_vs_chain}
\end{figure}

The same figures of merit were calculated for the DW2X ($n=30$) problem set, and can be seen in Figures \ref{DW2X_hist} and \ref{DW2X_median_success_vs_chain}. The advantages of $c=1.6$ are even more pronounced in these results than they were for the DW2 data.

\subsection{Performance of Parameter Setting Strategies}
Now we turn to a comparison of the different parameter setting strategies and how they performed over the two mixed SAT problem sets. If we look at the median success plots in Figures \ref{DW2_median_success_vs_chain} and  \ref{DW2X_median_success_vs_chain}, we see that the even, weighted, and weighted regularized strategies generally perform better than the single strategy, which declines rapidly with increasing chain coupling over the DW2 problem set, though the difference is not evident in the DW2X problem set. In order to understand these performance differences, we must remember that a choice of a parameter setting strategy is a choice of a time-dependent physical Hamiltonian with which quantum annealing will attempt to solve the optimization problem. The performance of QA depends critically on the characteristics of this physical Hamiltonian, especially in the time when the system approaches the minimum gap, but calculating the dynamics of the system at this point is more difficult than solving the original optimization problem. We can, however, readily examine the physical problem, which represents the Hamiltonian at the end of the QA evolution.

The figure of merit for the physical problem that is most relevant here is the minimum parameter distance (MPD), i.e., how close are the closest two values in the set $\{h_{i,k},J_{i,k;j,m},c\}$? If we take the first form of Equation \ref{eq_H_phys}, the values in this set are elements of the $n$ by $n$ coupling matrix $A$, and the MPD can be expressed simply as:

\begin{equation}
MPD=\min_{ijkl}{|A_{ij}-A_{kl}|}.
\end{equation}

\begin{figure}
\centering
\includegraphics[width=2.75in, angle=270]{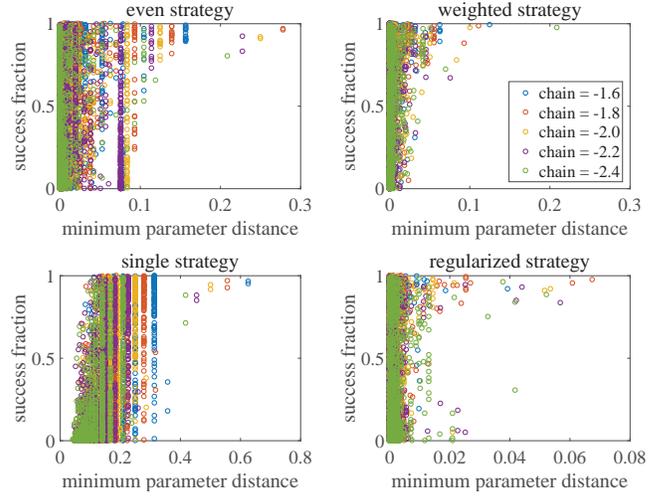}
\caption{Minimum parameter distance (MPD) of instances in DW2 problem set. One panel for each parameter setting strategy; each point represents a problem instance at a particular chain coupling. The single strategy exhibits the highest MPD overall due to the fact that logical problem terms aren't split up between multiple physical devices, but the MPD drops rapidly as the chain coupling increases because the logical value on a single physical device must be trimmed to accommodate chain couplings higher above $1$. The other three strategies show lower but more consistent MPD because they have multiple physical devices over which to distribute the logical values, blunting or negating entirely the impact of higher chain couplings.}
\label{precision_DW2}
\end{figure}

\begin{figure}
\centering
\includegraphics[width=2.75in, angle=270]{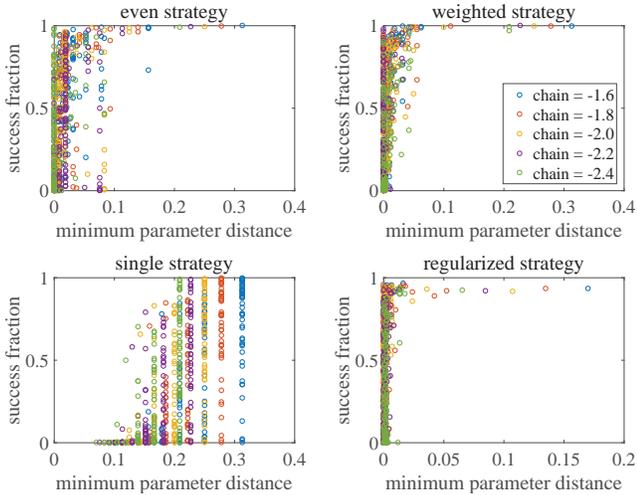}
\caption{Minimum parameter distance (MPD) of instances in DW2X problem set. One panel for each parameter setting strategy; each point represents a problem instance at a particular chain coupling. MPD values for all strategies are smaller here than in the DW2 problem set because the $n=30$ mixed SAT binary problems are larger and more complicated. Although the DW2X chip can handle lower MPD than the DW2 chip, the MPD demands of this problem set are such that all strategies suffer as chain coupling increases.}
\label{precision_DW2X}
\end{figure}

The closer these physical parameters are, the more likely it is that they will cross due to one of the many sources of noise on-chip. A parameter setting with more separation between all physical values, then, is preferable to one that drives two or more values very close together (i.e. we prefer larger MPD because it signifies a physical problem that is more robust to noise). Figures \ref{precision_DW2} and \ref{precision_DW2X} show what is happening with the MPD of the instances in the problem sets when they are parameterized with different strategies and chain couplings.

\subsection{Effect of Spin Reversal Transformations}
Spin reversal transformations (SRTs) are another way to change the physical problem representation. The transformation is represented by a reversal vector $r\in\{-1,+1\}^N$, which has length $N$ (the number of physical qubits) and creates a transformed $H_{physical}$

\begin{align}
H'_{physical} = & \sum_{i=1}^{n} \sum_{k=1}^{K_i} r_{i,k}h_{i,k} s_{i,k} + \\ & \sum_{(i,j)\in E_{l}} \sum_{(i,k;j,m)\in E_p} r_{i,k}r_{j,m}J_{i,k;j,m} s_{i,k} s_{j,m} + H_{chain}
\end{align}

\begin{figure}
\centering
\includegraphics[width=2.75in]{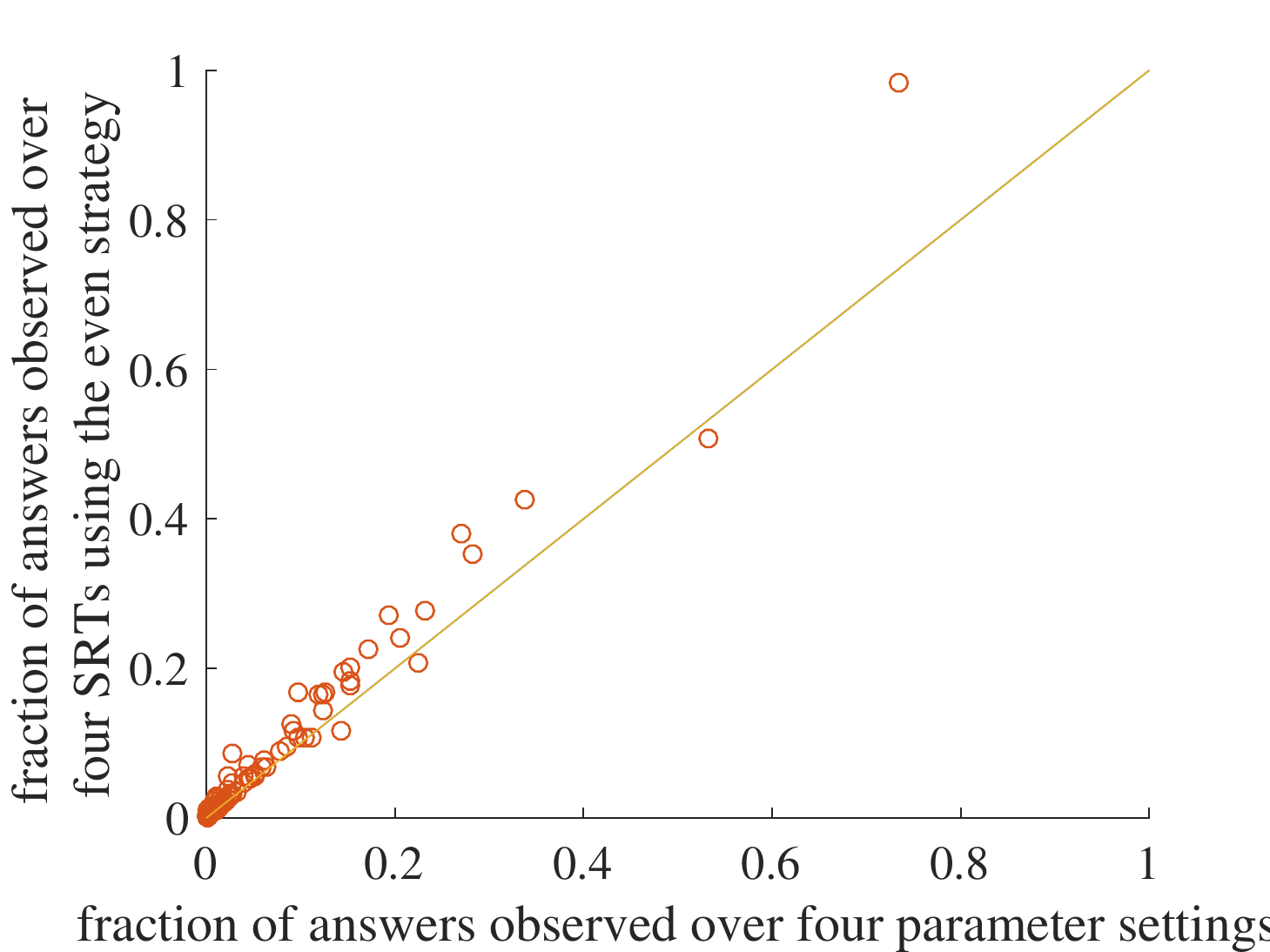}
\caption{Fraction of answer set observed using four spin reversal transformations vs. four parameter setting strategies. Most instances fall on or above the break-even line, yielding more unique answers when subjected to a spin reversal transformation than when re-parameterized.}
\label{frac_gauges_vs_params}
\end{figure}

\begin{figure}
\centering
\includegraphics[width=2.75in, angle=270]{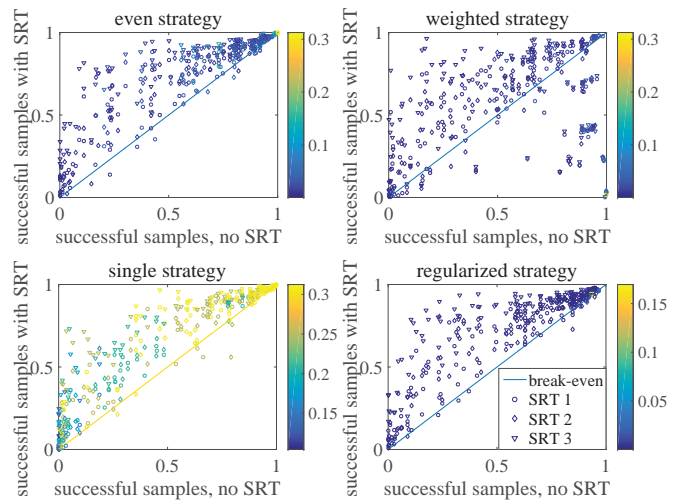}
\caption{Success probability of DW2X problem set after spin reversal transformations (SRTs). One panel for each parameter setting strategy. For an overwhelming majority of instances, all spin reversal transformations studied yield an enhancement in success probability over the identity. The color scale indicates MPD, which is more strongly associated with a higher overall success probability than with improvements from any particular SRT.}
\label{success_gauges_vs_params}
\end{figure}

The use of multiple parameter settings or spin reversal transformations becomes important when we consider sampling from a large solution space. By changing the physical problem, we may gain access to different parts of the solution space, allowing us to see a wider variety of answers. The use of SRTs to boost success probability by averaging out noise is well established using the argument that SRTs can flip the effect of persistent bias on individual qubits and couplers \cite{ronnow2014defining,Perdomo-Ortiz2015}. If not averaged out, these biases can push the system in different directions depending on the effect of the current SRT. We subjected the DW2X problem set to four spin reversal transformations (the same four used in \cite{adachi2015application}, including the identity transformation) to determine whether multiple parameter settings or SRTs provide more unique answers and better success probability. The results, shown in Figures \ref{frac_gauges_vs_params} and \ref{success_gauges_vs_params}, favor the spin reversal transformations for this purpose.

\section{Conclusions}
\label{conclusions}
The performance of parameter setting strategies on D-Wave quantum annealers is affected strongly by the MPD of the physical problem generated. This is good news because physical problem MPD can be calculated efficiently for problems of application scale, as opposed to physically important but computationally inaccessible values like the minimum gap of the time-dependent quantum annealing energy spectrum. Distribution of logical bias and coupling values over more than one physical device is also a desirable characteristic of a parameter setting strategy because it avoids single qubit failure modes and makes the success of the calculation less dependent on the choice of chain coupling. The even and single parameter setting strategies described here are recommended to QA programmers in the field for their ease of implementation and favorable characteristics. The even strategy achieves distribution of logical values, but the single strategy may be appropriate for logical problems with unfavorable MPD \textit{a priori}. New parameter setting strategies with favorable logical value distribution and physical problem MPD may still emerge from future research.

For this problem set, we found that spin reversal transformations performed better than alternate parameter setting strategies to boost success probability and solution sampling diversity. Reasons for this may include the similarity of three out of the four parameter setting strategies (even, weighted, and weighted regularized); these strategies may have produced physical problems that were more meaningfully consistent than the same parameter setting with a spin reversal transformation. Additionally, the parameter setting strategies may have suffered from being studied with no spin reversal transformation applied; for this chip and problem set, the identity transformed problems exhibited the poorest performance. Whatever the reason, the utility of spin reversal transformations is a positive result because they are easy to generate and lightweight to perform, rendering them a good tool for practical QA programmers.



\begin{thebibliography}{10}
\providecommand{\url}[1]{#1}
\csname url@samestyle\endcsname
\providecommand{\newblock}{\relax}
\providecommand{\bibinfo}[2]{#2}
\providecommand{\BIBentrySTDinterwordspacing}{\spaceskip=0pt\relax}
\providecommand{\BIBentryALTinterwordstretchfactor}{4}
\providecommand{\BIBentryALTinterwordspacing}{\spaceskip=\fontdimen2\font plus
\BIBentryALTinterwordstretchfactor\fontdimen3\font minus
  \fontdimen4\font\relax}
\providecommand{\BIBforeignlanguage}[2]{{%
\expandafter\ifx\csname l@#1\endcsname\relax
\typeout{** WARNING: IEEEtran.bst: No hyphenation pattern has been}%
\typeout{** loaded for the language `#1'. Using the pattern for}%
\typeout{** the default language instead.}%
\else
\language=\csname l@#1\endcsname
\fi
#2}}
\providecommand{\BIBdecl}{\relax}
\BIBdecl

\bibitem{hen2015probing}
I.~Hen, J.~Job, T.~Albash, T.~F. R{\o}nnow, M.~Troyer, and D.~A. Lidar,
  ``Probing for quantum speedup in spin-glass problems with planted
  solutions,'' \emph{Physical Review A}, vol.~92, no.~4, p. 042325, 2015.

\bibitem{king2015performance}
A.~D. King, ``Performance of a quantum annealer on range-limited constraint
  satisfaction problems,'' \emph{arXiv preprint arXiv:1502.02098}, 2015.

\bibitem{ronnow2014defining}
T.~F. R{\o}nnow, Z.~Wang, J.~Job, S.~Boixo, S.~V. Isakov, D.~Wecker, J.~M.
  Martinis, D.~A. Lidar, and M.~Troyer, ``Defining and detecting quantum
  speedup,'' \emph{Science}, vol. 345, no. 6195, pp. 420--424, 2014.

\bibitem{boixo2014evidence}
S.~Boixo, T.~F. R{\o}nnow, S.~V. Isakov, Z.~Wang, D.~Wecker, D.~A. Lidar, J.~M.
  Martinis, and M.~Troyer, ``Evidence for quantum annealing with more than one
  hundred qubits,'' \emph{Nature Physics}, vol.~10, no.~3, pp. 218--224, 2014.

\bibitem{King2014}
A.~D. King and C.~C. McGeoch, ``Algorithm engineering for a quantum annealing
  platform,'' \emph{arxiv}, October 2014.

\bibitem{hen2016quantum}
I.~Hen and F.~M. Spedalieri, ``Quantum annealing for constrained
  optimization,'' \emph{Physical Review Applied}, vol.~5, no.~3, p. 034007,
  2016.

\bibitem{rieffel2015case}
E.~G. Rieffel, D.~Venturelli, B.~O�Gorman, M.~B. Do, E.~M. Prystay, and V.~N.
  Smelyanskiy, ``A case study in programming a quantum annealer for hard
  operational planning problems,'' \emph{Quantum Information Processing},
  vol.~14, no.~1, pp. 1--36, 2015.

\bibitem{venturelli2015jobshop}
D.~Venturelli, D.~J. Marchand, and G.~Rojo, ``Quantum annealing implementation
  of job-shop scheduling,'' \emph{arXiv preprint arXiv:1506.08479}, 2015.

\bibitem{perdomo2015fault}
A.~Perdomo-Ortiz, J.~Fluegemann, S.~Narasimhan, R.~Biswas, and V.~N.
  Smelyanskiy, ``A quantum annealing approach for fault detection and diagnosis
  of graph-based systems,'' \emph{The European Physical Journal Special
  Topics}, vol. 224, no.~1, pp. 131--148, 2015.

\bibitem{Zick2015}
K.~M. Zick, O.~Shehab, and M.~French, ``Experimental quantum annealing: case
  study involving the graph isomorphism problem,'' \emph{arxiv}, March 2015.

\bibitem{adachi2015application}
S.~H. Adachi and M.~P. Henderson, ``Application of quantum annealing to
  training of deep neural networks,'' \emph{arXiv preprint arXiv:1510.06356},
  2015.

\bibitem{neven2009nips}
H.~Neven, V.~S. Denchev, M.~Drew-Brook, J.~Zhang, W.~G. Macready, and G.~Rose,
  ``Nips 2009 demonstration: Binary classification using hardware
  implementation of quantum annealing,'' \emph{Quantum}, pp. 1--17, 2009.

\bibitem{pudenz2013quantum}
K.~L. Pudenz and D.~A. Lidar, ``Quantum adiabatic machine learning,''
  \emph{Quantum information processing}, vol.~12, no.~5, pp. 2027--2070, 2013.

\bibitem{Perdomo-Ortiz2015}
A.~Perdomo-Ortiz, J.~Fluegemann, R.~Biswas, and V.~N. Smelyanskiy, ``A
  performance estimator for quantum annealers: Gauge selection and parameter
  setting,'' \emph{arxiv}, March 2015.

\bibitem{Perdomo-Ortiz2015a}
A.~Perdomo-Ortiz, B.~O'Gorman, J.~Fluegemann, R.~Biswas, and V.~N. Smelyanskiy,
  ``Determination and correction of persistent biases in quantum annealers,''
  \emph{arxiv}, March 2015.

\bibitem{venturelli2015quantum}
D.~Venturelli, S.~Mandr{\`a}, S.~Knysh, B.~O�Gorman, R.~Biswas, and
  V.~Smelyanskiy, ``Quantum optimization of fully connected spin glasses,''
  \emph{Physical Review X}, vol.~5, no.~3, p. 031040, 2015.

\bibitem{cai2014practical}
J.~Cai, W.~G. Macready, and A.~Roy, ``A practical heuristic for finding graph
  minors,'' \emph{arXiv preprint arXiv:1406.2741}, 2014.

\bibitem{hoos2000satllb}
H.~Hoos and T.~Stiitzle, ``Satllb: An online resource for research on sat,''
  \emph{Sat2000: highlights of satisfiability research in the year 2000}, vol.
  283, 2000.

\bibitem{toda2015implementing}
T.~Toda and T.~Soh, ``Implementing efficient all solutions sat solvers,''
  \emph{arXiv preprint arXiv:1510.00523}, 2015.

\bibitem{johnson2011quantum}
M.~Johnson, M.~Amin, S.~Gildert, T.~Lanting, F.~Hamze, N.~Dickson, R.~Harris,
  A.~Berkley, J.~Johansson, P.~Bunyk \emph{et~al.}, ``Quantum annealing with
  manufactured spins,'' \emph{Nature}, vol. 473, no. 7346, pp. 194--198, 2011.

\bibitem{lanting2014entanglement}
T.~Lanting, A.~Przybysz, A.~Y. Smirnov, F.~Spedalieri, M.~Amin, A.~Berkley,
  R.~Harris, F.~Altomare, S.~Boixo, P.~Bunyk \emph{et~al.}, ``Entanglement in a
  quantum annealing processor,'' \emph{Physical Review X}, vol.~4, no.~2, p.
  021041, 2014.

\bibitem{bunyk2014architectural}
P.~I. Bunyk, E.~M. Hoskinson, M.~W. Johnson, E.~Tolkacheva, F.~Altomare, A.~J.
  Berkley, R.~Harris, J.~P. Hilton, T.~Lanting, A.~J. Przybysz \emph{et~al.},
  ``Architectural considerations in the design of a superconducting quantum
  annealing processor,'' \emph{Applied Superconductivity, IEEE Transactions
  on}, vol.~24, no.~4, pp. 1--10, 2014.

\end{thebibliography}
\end{document}